\documentclass[journal]{IEEEtran}

\usepackage{amsmath}   
\usepackage{amssymb}

\newtheorem{thm}{Theorem}
\newtheorem{prop}{Proposition}
\newtheorem{cor}{Corollary}
\newtheorem{lem}{Lemma}
\newtheorem{defn}{Definition}
\newtheorem{fact}{Fact}
\newtheorem{rmk}{Remark}

\newcommand{\bc}{\begin{center}}
\newcommand{\ec}{\end{center}}
\newcommand{\bt}{\begin{tabular}}
\newcommand{\et}{\end{tabular}} 
\newcommand{\bea}{\begin{eqnarray}}
\newcommand{\eea}{\end{eqnarray}}

\newcommand{\ba}{\begin{array}}
\newcommand{\ea}{\end{array}}

\def\be{\begin{eqnarray}}
\def\ee{\end{eqnarray}}
\def\ben{\begin{eqnarray*}}
\def\een{\end{eqnarray*}}

\newcommand{\ra} {\rightarrow}

\newcommand{\Ra} {\Rightarrow}

\newcommand{\eqd}{\mbox{$ \;\stackrel{(d)}{=}\; $}}

\newcommand{\nth}{\frac{1}{n}}

\newcommand{\RL}{{\mathbb R}}

\newcommand{\calH}{\mbox{${\cal H}$}}

\newcommand{\Nat}{\mathbb{N}}
\newcommand{\Sumn}{\sum_{i=1}^{n}}

\newcommand{\dou}{\partial}

\newcommand{\VAR}{\mbox{\rm Var}}

\newcommand{\lam}{\lambda}

\def\sq{$\Box$}

\def\qed{\ifmmode\sq\else{\unskip\nobreak\hfil
\penalty50\hskip1em\null\nobreak\hfil\sq
\parfillskip=0pt\finalhyphendemerits=0\endgraf}\fi\par\medbreak}

\newsavebox{\junk}
\savebox{\junk}[1.6mm]{\hbox{$|\!|\!|$}}

\def\half{{\mathchoice{\textstyle \frac{1}{2}}%
{\frac{1}{2}}%
{\hbox{\tiny $\frac{1}{2}$}}%
{\hbox{\tiny $\frac{1}{2}$}} }}

 \def\eq#1/{(\ref{#1})}

\def\eq#1/{(\ref{e:#1})}

\newcommand{\setS}{\mbox{${\bf s}$}}
\newcommand{\sumS}{\sum_{\setS\in\collS}}
\newcommand{\setT}{\mbox{${\bf t}$}}
\newcommand{\collS}{\mathcal{C}}
\newcommand{\Etb}{\bar{E}_{\setT}}
\newcommand{\asb}{\bar{a}_{\setS}}
\newcommand{\ws}{w_{\setS}}
\newcommand{\mus}{\mu_{\setS}}
\newcommand{\bs}{\beta_{\setS}}

\newcommand{\rth}{\frac{1}{r}}
\newcommand{\calN}{\mathcal{N}}

\begin{document}
%
\title{Generalized Entropy Power Inequalities and Monotonicity Properties of Information}
\author{Mokshay~Madiman,~\IEEEmembership{Member,~IEEE,}
        and~Andrew~Barron,~\IEEEmembership{Senior Member,~IEEE}
\thanks{Manuscript received April 15, 2006; revised March 16, 2007.
The material in this paper was presented in part at the 
Inaugural Workshop, Information Theory and Applications Center, 
San Diego, CA, February 2006, and at the IEEE International 
Symposium on Information Theory, Seattle, WA, July 2006.}
\thanks{The authors are with 
the Department of Statistics, Yale University,
24 Hillhouse Avenue, New Haven, CT 06511, USA.
Email: {\tt mokshay.madiman@yale.edu, andrew.barron@yale.edu}}
\thanks{Communicated by V. A. Vaishampayan, Associate Editor at Large.}}
%
%
%
\markboth{To appear in IEEE Transactions on Information Theory, Vol. 53, No. 7, July 2007}{Madiman and Barron \MakeLowercase
: Generalized Entropy Power Inequalities}
%



\maketitle

\begin{abstract}
New families of Fisher information and entropy power inequalities
for sums of independent random variables are presented. These inequalities relate the information
in the sum of $n$ independent random variables to the information contained 
in sums over subsets of the random variables, for an arbitrary collection of subsets.
As a consequence, a simple proof of the monotonicity of information
in central limit theorems is obtained, both in the setting of  i.i.d. summands as well
as in the more general setting of independent summands with variance-standardized sums.
\end{abstract}


%

\section{Introduction}
\label{sec:intro}

\PARstart{L}{et}  $X_1, X_2, \ldots,X_{n}$ be independent
random variables with densities and finite variances. Let $H$
denote the (differential) entropy, i.e., 
if $f$ is the probability density function of $X$, then
$H(X)= -E[\log f(X)]$. The classical entropy power inequality
of Shannon \cite{Sha48} and Stam \cite{Sta59} states
\be\label{sha-epi}
e^{2H(X_{1}+\ldots+X_{n})} \geq \sum_{j=1}^{n} e^{2H(X_{j})} .
\ee
In 2004, Artstein, Ball, Barthe and Naor \cite{ABBN04:1} (hereafter denoted by
ABBN \cite{ABBN04:1}) proved
a new entropy power inequality
\be\label{abbn-epi}
e^{2H(X_{1}+\ldots+X_{n})} \geq \frac{1}{n-1} \sum_{i=1}^{n} e^{2H \big( \sum_{j\neq i} X_{j}\big)  } ,
\ee
where each term involves the entropy of the sum of $n-1$ of the variables 
excluding the $i$-th, and presented its implications for the
monotonicity of entropy in the central limit theorem. 
It is not hard to see, by repeated application of \eqref{abbn-epi} 
for a succession of values of $n$, that \eqref{abbn-epi} in fact implies
the inequality \eqref{m-epi} and hence 
\eqref{sha-epi}. 
We will present below a generalized entropy power inequality for subset sums that 
subsumes both \eqref{abbn-epi} and \eqref{sha-epi} and also implies
several other interesting inequalities. We provide simple and 
easily interpretable proofs of all of these inequalities. In particular,
this provides a simplified understanding of the monotonicity of entropy in  
central limit theorems. A similar independent and contemporaneous development
of the monotonicity of entropy is given by Tulino and Verd\'u \cite{TV06}.

Our generalized entropy power inequality for subset sums is as follows: 
if $\collS$ is an arbitrary collection of subsets of
$\{1,2,\ldots, n\}$, then
\be\label{our-epi}
e^{2H(X_{1}+\ldots+X_{n})} \geq \frac{1}{{r}} \sumS e^{ 2H \big( \sum_{j\in\setS} X_{j}\big) } ,
\ee
where ${r}$ is the maximum number of sets in $\collS$ in which any one index appears.
In particular, note that
\begin{enumerate}
\item Choosing $\collS$ to be the class $\collS_{1}$ of all singletons
yields $r=1$ and hence \eqref{sha-epi}.
\item Choosing $\collS$ to be the class $\collS_{n-1}$ of all sets of $n-1$ elements
yields $r=n-1$ and hence \eqref{abbn-epi}.
\item Choosing $\collS$ to be the class $\collS_{m}$ of all sets of $m$ elements
yields $r=\binom{n-1}{m-1}$ and hence  the inequality
\be\label{m-epi}\begin{split}
\exp\bigg\{ & 2H\bigg({X_1+\ldots+X_n}\bigg) \bigg\} \geq \\
&\frac{1}{\binom{n-1}{m-1}} \sum_{\setS\in\collS_{m}} 
\exp\bigg\{2H\bigg(\sum_{i\in\setS} X_{i}\bigg) \bigg\} .
\end{split}\ee
\item Choosing $\collS$ to be the class of all sets of $k$
consecutive integers yields $r=\min{\{k,n+1-k\}}$ and hence the inequality
\ben
\begin{split}
\exp\bigg\{ 2 & H\big(X_1+\ldots+X_n\big) \bigg\} \geq \\
&\frac{1}{\min{\{k,n+1-k\}}} \sumS \exp\bigg\{2H\bigg(\sum_{i\in\setS} X_{i}\bigg) \bigg\} .
\end{split}\een
\end{enumerate}
In general, the inequality \eqref{our-epi} clearly yields a whole family 
of entropy power inequalities, for arbitrary collections of subsets. 
Furthermore, equality holds in any of these inequalities if and only if 
the $X_{i}$ are normally distributed and the collection $\collS$ is
``nice'' in a sense that will be made precise later. 

These inequalities are relevant for the examination of monotonicity in
central limit theorems. Indeed, if $X_{1}$ and $X_{2}$ are independent
and identically distributed (i.i.d.), then \eqref{sha-epi} is equivalent to
\be\label{step2}
H\bigg(\frac{X_1+X_2}{\sqrt{2}}\bigg) \geq H(X_1) ,
\ee
by using the scaling $H(aX)=H(X)+\log |a|$.
This fact implies that the entropy of the standardized sums 
$Y_{n}=\frac{\sum_{i=1}^{n} X_{i}}{\sqrt{n}}$ for i.i.d. $X_{i}$ 
increases along the powers-of-2 
subsequence, i.e., $H(Y_{2^k})$ is non-decreasing in $k$. 
Characterization of the change in information quantities on
doubling of sample size was used in proofs of central limit theorems
by Shimizu \cite{Shi75}, Brown \cite{Bro82}, Barron \cite{Bar86},
Carlen and Soffer \cite{CS91}, Johnson and Barron \cite{JB04}, and
Artstein, Ball, Barthe and Naor \cite{ABBN04:2}. In particular, Barron \cite{Bar86}
showed that the sequence $\{H(Y_{n})\}$ 
converges to the entropy of the normal;
this, incidentally, is equivalent to the convergence to 0 of the
relative entropy (Kullback divergence) from a normal distribution.
ABBN \cite{ABBN04:1}  showed  that $H(Y_n)$ is in fact a  
non-decreasing sequence for every $n$, solving a long-standing conjecture.
In fact, \eqref{abbn-epi} is equivalent in the i.i.d. case to
the monotonicity property
\be\label{stepn}
H\bigg(\frac{X_1+\ldots+X_n}{\sqrt{n}}\bigg) \geq H\bigg(\frac{X_1+\ldots+X_{n-1}}{\sqrt{n-1}}\bigg) .
\ee
Note that the presence of the factor $n-1$ (rather than $n$) in the denominator of \eqref{abbn-epi}
is crucial for this monotonicity. 

Likewise, for sums of independent
random variables, the inequality \eqref{m-epi} is equivalent to 
``monotonicity on average'' properties for certain standardizations; for instance, 
\ben
H\bigg(\frac{X_1+\ldots+X_n}{\sqrt{n}}\bigg)  \geq 
\frac{1}{\binom{n}{m}} \sum_{\setS\in\collS_{m}} H\bigg(\frac{\sum_{i\in\setS} X_{i}}{\sqrt{m}}\bigg)  .
\een
A similar monotonicity also holds, as we shall show, for arbitrary collections $\collS$ 
and even when the sums are standardized by their variances.
Here again the factor ${r}$ (rather than the cardinality $|\collS|$ of the
collection) in the denominator of  \eqref{our-epi} for the unstandardized version is crucial.

\vspace{.2in}
\noindent{\bf Outline of our development.}
\vspace{.2in}

We find that the main inequality \eqref{our-epi} (and hence all of the above inequalities)
as well as corresponding inequalities for Fisher information can be be proved by
simple tools. 
Two of these tools, a convolution identity for
score functions 
and the relationship between Fisher information and entropy
(discussed in Section~\ref{sec:tools}),
are familiar in past work on information inequalities.
An additional trick is needed to obtain the denominator ${r}$ in 
\eqref{our-epi}. This is a simple variance drop inequality for statistics expressible via 
sums of functions of subsets of a collection of variables, particular cases of which are familiar 
in other statistical contexts (as we shall discuss). 

We recall that for a random variable $X$ with differentiable density $f$,
the score function is $\rho(x)=\frac{\dou}{\dou x}\log f(x)$,
the score is the random variable $\rho(X)$,
and its Fisher information is $I(X)=E[\rho^{2}(X)]$.

Suppose for the consideration of Fisher information that the independent
random variables $X_{1},\ldots, X_{n}$ have absolutely continuous densities.
To outline the heart of the matter, the first step boils down to the geometry
of projections (conditional expectations). Let $\rho_{tot}$ be the score
of the total sum $\sum_{i=1}^{n}X_{i}$ and let $\rho_{\setS}$
be the score of the subset sum $\sum_{i\in\setS}X_{i}$.
As we recall in Section~\ref{sec:tools}, 
$ \rho_{tot}$   
is the conditional expectation (or $L^{2}$ projection) of each of these subset 
sum scores given the total sum. 
Consequently, any convex combination
$\sumS\ws \rho_{\setS}$ also has projection
\ben 
\rho_{tot}=E\bigg[\sumS\ws\rho_{\setS}\bigg| \sum_{i=1}^{n}X_{i}\bigg]
\een
and the Fisher information $I(X_1+\ldots+X_n)=E[\rho_{tot}^{2}]$
has the bound
\be\label{proj-ol}
E[\rho_{tot}^{2}] \leq E\bigg[ \bigg(\sumS\ws\rho_{\setS}\bigg)^{2} \bigg] .
\ee
For non-overlapping subsets, the independence and zero mean properties of the
scores provide a direct means to express the right side in terms
of the Fisher informations of the subset sums
(yielding the traditional Blachman \cite{Bla65} proof of Stam's inequality
for Fisher information). In contrast, the case of overlapping subsets requires fresh 
consideration. Whereas a naive application of Cauchy-Schwarz would yield
a loose bound of 
$|\collS| \sumS\ws^{2} E\rho_{\setS}^{2}$,
instead a variance drop lemma yields that the right side of \eqref{proj-ol}
is not more than
$r \sumS\ws^{2} E\rho_{\setS}^{2}$
if each $i$ is in at most $r$ subsets of $\collS$.
Consequently,
\be\label{central-I}
I(X_1+\ldots+X_n) \leq  {r} \sumS 
\ws^{2} I\bigg( \sum_{i\in\setS} X_{i} \bigg) 
\ee
for any 
weights $w_{\setS}$ that add to 1 over all subsets
$\setS\subset\{1,\ldots,n\}$ in $\collS$. See Sections~\ref{sec:tools} 
and \ref{sec:fii} for details.
Optimizing over $w$ yields an inequality for inverse Fisher information 
that extends the Fisher information inequalities of Stam and ABBN:
\be\label{our-fii}
\frac{1}{I(X_{1}+\ldots +X_{n})} \geq \frac{1}{{r}} 
\sumS \frac{1}{I( \sum_{i\in\setS} X_{i})}  .
\ee

Alternatively, using a scaling property of Fisher information to
re-express our core inequality \eqref{central-I}, we see that
the Fisher information of the total sum is bounded by a convex combination
of  Fisher informations of scaled subset sums:
\be\label{central-H}
I(X_1+\ldots+X_n) \leq  \sumS 
\ws I\bigg( \frac{\sum_{i\in\setS} X_{i}}{\sqrt{\ws {r}}} \bigg) .
\ee
This integrates to give an inequality for entropy that is an extension
of the ``linear form of the entropy power inequality'' 
developed by Dembo et al \cite{DCT91}.
Specifically we obtain 
\be\label{strand-H}
H(X_1+\ldots+X_n) \leq  \sumS 
\ws H\bigg( \frac{\sum_{i\in\setS} X_{i}}{\sqrt{\ws {r}}} \bigg) .
\ee
See Section~\ref{sec:ent} for details.
Likewise using the scaling property of entropy on \eqref{strand-H} and optimizing over
$w$ yields our extension \eqref{our-epi} of the entropy power inequality, 
described in Section~\ref{sec:epi}.

To relate this chain of ideas to other recent work, 
we point out that ABBN \cite{ABBN04:1}
was the first to show use of a variance drop lemma
for information inequality development
(in the leave-one-out case of the collection $\collS_{n-1}$).
For that case what is new in our presentation is going straight to
the projection inequality \eqref{proj-ol} followed by
the variance drop inequality, bypassing any need for the
elaborate variational characterization of Fisher information that is
an essential ingredient of ABBN \cite{ABBN04:1}.
Moreover, whereas ABBN \cite{ABBN04:1} requires that 
the random variables have a $C^{2}$ density for monotonicity of Fisher
information, absolute continuity of the density (as minimally needed to define the score function)
suffices in our approach. Independently and contemporaneously to
our work, Tulino and Verd\'u \cite{TV06} also found a similar simplified
proof of monotonicity properties of entropy and entropy derivative with a
clever interpretation (its relationship to our approach is described in Section~\ref{sec:iid}
after we complete the presentation of our development).
Furthermore, a recent preprint of Shlyakhtenko \cite{Shl05:pre}
proves an analogue of the information inequalities in the leave-one-out case 
for non-commutative or ``free'' probability theory. 
In that manuscript, he also gives a proof for the classical setting
assuming finiteness of all moments, whereas our direct proof requires only finite 
variance. Our proof also reveals in a simple manner the cases of equality in \eqref{stepn},
for which an alternative approach in the free probability setting is in Schultz \cite{Sch05:pre}.

While Section~\ref{sec:iid} gives a direct proof of the
monotonicity of entropy in the central limit theorem for 
i.i.d. summands, Section~\ref{sec:avg} applies the preceding
inequalities to study sums of non-identically distributed random 
variables under appropriate scalings. In particular, we show 
that ``entropy is monotone on average'' in the setting
of variance-standardized sums.

Our subset sum inequalities are tight (with equality in the Gaussian case)
for balanced collections of subsets, as will be defined in Section~\ref{sec:tools}. 
In Section~\ref{sec:ref}, we present refined versions of
our inequalities that can even be tight for certain unbalanced collections.

Section~\ref{sec:concl} concludes with some discussion on potential directions 
of application of our results and methods. In particular, beyond the 
connection with central limit theorems, we also discuss potential connections of
our results with distributed statistical estimation, graph theory and 
multi-user information theory.

\vspace{.2in}
\noindent{\bf Form of the inequalities.}
\vspace{.2in}

If $\psi(X)$ represents either the inverse Fisher information or the entropy 
power of $X$, then our inequalities above take the form
\be\label{form}
{r} \psi \big(X_1+\ldots+X_n\big) 
\geq  \sumS \psi \bigg(\sum_{i\in \setS} X_{i}\bigg) .
\ee
We motivate the form \eqref{form} 
using the following almost trivial fact. 

\par\vspace{.15in}
\noindent\begin{fact}\label{f:add}
For arbitrary numbers $\{a_{i}:i=1,2,\ldots,n\}$, 
\be
\sumS \sum_{i\in \setS} a_{i}
= {r} \sum_{i=1}^{n} a_{i} ,
\ee
if each index $i$ appears in $\collS$ the same number of times $r$.
\end{fact}
\par\vspace{.15in}

Indeed,
\ben\begin{split}
\sumS \sum_{i\in \setS} a_{i}
&= \sum_{i=1}^{n} \sum_{\setS\ni i, \setS\in\collS} a_{i}\\
&= \sum_{i=1}^{n} {r} a_{i}
= {r} \sum_{i=1}^{n} a_{i}.
\end{split}\een

If Fact~\ref{f:add} is thought of as $\collS$-additivity
of the sum function for real numbers,
then \eqref{our-fii} and \eqref{our-epi}
represent the $\collS$-{\it superadditivity} of
inverse Fisher information and entropy power functionals 
respectively with respect to convolution of the arguments. 
In the case of normal random variables, the
inverse Fisher information and the entropy power equal
the variance. Thus in that case 
 \eqref{our-fii} and \eqref{our-epi}
 become Fact~\ref{f:add} with $a_{i}$
 equal to the variance of $X_{i}$.

\section{Score Functions and Projections}
\label{sec:tools}

We use  $\rho(x)=\frac{f'(x)}{f(x)}$ to denote the
(almost everywhere defined) score function of the random variable $X$ with absolutely 
continuous probability density function $f$. 
The score $\rho(X)$ has zero mean, and its 
variance is just the Fisher information $I(X)$.

The first tool we need is a projection property of score functions of sums of independent random variables,
which is well-known for smooth densities (cf., Blachman \cite{Bla65}).
For completeness, we give the proof. As shown by Johnson and Barron \cite{JB04}, 
it is sufficient that the densities are absolutely continuous;
see \cite{JB04}[Appendix 1] for an explanation of why this is so.

\par\vspace{.15in}
\noindent\begin{lem}[{\sc Convolution identity for scores}]\label{lem:conv}
If $V_{1}$ and $V_{2}$ are independent random variables, 
and $V_{1}$ has an absolutely continuous density with
score $\rho_{1}$, then $V_{1}+V_{2}$ has the score function
\be
\rho(v)= E [\rho_{{1}}(V_{1})| V_{1}+V_{2}=v]
\ee
\end{lem}

\par\vspace{.15in}
\noindent\begin{proof}
Let $f_{{1}}$ and $f$ be the densities of
$V_{1}$ and $V=V_{1}+V_{2}$ respectively. Then, either bringing the 
derivative inside the integral for the smooth case, or via the
more general formalism in \cite{JB04},
\ben\begin{split}
f'(v)&=\frac{\dou}{\dou v} E[ f_{1}(v-V_{2}) ] \\
&= E [  f_{{1}}'(v-V_{2}) ] \\
&= E [  f_{{1}}(v-V_{2}) \rho_{1}(v-V_{2})]
\end{split}\een
so that
\ben\begin{split}
\rho(v)&=\frac{f'(v)}{f(v)} \\
&= E \bigg[  \frac{f_{{1}}(v-V_{2})}{f(v)} \rho_{1}(v-V_{2}) \bigg] \\
&= E [\rho_{1}(V_{1})| V_{1}+V_{2}=v].
\end{split}\een
\end{proof}
\par\vspace{.15in}

The second tool we need is a ``variance drop lemma'', the history of which
we discuss in remarks after the proof below.
The following conventions are useful: 
\begin{itemize}
\item 
$[n]$ is the index set $\{1,2,\ldots,n\}$. 
\item 
For any $\setS\subset [n]$, 
$X_{\setS}$ stands for the collection of random
variables $(X_{i}:i\in \setS)$, with the indices taken
in their natural (increasing) order.
\item 
For $\psi_{\setS}:\RL^{|\setS|}\ra\RL$, we write 
$\psi_{\setS}(x_{\setS})$ for a function of $x_{\setS}$
for any $\setS\subset[n]$, so that
$\psi_{\setS}(x_{\setS})\equiv\psi_{\setS}(x_{k_{1}},\ldots,x_{k_{|\setS|}})$,
where $k_{1}<k_{2}<\ldots < k_{|\setS|}$ are the ordered indices in $\setS$.
\item 
We say that a function $U:\RL^{n}\ra\RL$ is {\it $\collS$-additive}
if it can be expressed in the form
$\sumS \psi_{\setS}(x_{\setS})$.
\end{itemize}

The following notions are not required for the inequalities we present,
but help to clarify the cases of equality. 
\begin{itemize}
\item 
A collection $\collS$ of subsets of $[n]$ is said to be {\it discriminating} if 
for any distinct indices $i$ and $j$ in $[n]$, there is a set in $\collS$ that
contains $i$ but not $j$. Note that all the collections introduced in Section~\ref{sec:intro}
were discriminating. 
\item
A collection $\collS$ of subsets of $[n]$ is said to be {\it balanced} if 
each index $i$ in $[n]$ appears in the same number (namely, ${r}$)
of sets in $\collS$. 
\item 
A function $f:\RL^{d}\ra\RL$ is {\it additive}
if there exist functions $f_{i}:\RL\ra\RL$ such that
$f(x_{1},\ldots,x_{d})=\sum_{i=1}^{d} f_{i}(x_{i})$, i.e.,
if it is $\collS_{1}$-additive.
\end{itemize}

\par\vspace{.15in}
\noindent\begin{lem}[{\sc Variance drop}]\label{lem:vardrop}
Let 
$U(X_{1},\ldots,X_{n})= \sumS \psi_{\setS}(X_{\setS}) $
be a $\collS$-additive function with 
mean zero components, i.e., $E\psi_{\setS}(X_{\setS})=0$
for each $\setS\in\collS$.
Then
\be
EU^{2} \leq {r} \sumS 
E\big\{\psi_{\setS}(X_{\setS})\big\}^{2} ,
\ee
where $r$ is the maximum number of subsets $\setS\in\collS$
in which any index appears.
When $\collS$ is a discriminating collection, equality can hold 
only if each $\psi_{\setS}$ is an additive function.
\end{lem}

\par\vspace{.15in}
\noindent\begin{proof}
For every subset $\setT$ of $[n]$, let $\Etb$ 
be the ANOVA projection onto the space of functions
of $X_{\setT}$ (see the Appendix for details). By performing 
the ANOVA decomposition on each $\psi_{\setS}$,
we have
\be\begin{split}
EU^{2} 
&= E\bigg( \sumS 
\sum_{\setT\subset \setS} \Etb \psi_{\setS}(X_{\setS}) \bigg)^{2} \\
&= E\bigg( \sum_{\setT} \Etb
\sum_{\setS\supset \setT, \setS\in\collS}   \psi_{\setS}(X_{\setS}) \bigg)^{2} \\
&=  \sum_{\setT} E\bigg( 
\sum_{\setS\supset \setT, \setS\in\collS}  \Etb\psi_{\setS}(X_{\setS}) \bigg)^{2} ,
\end{split}\ee
using the orthogonality of the ANOVA decomposition of $U$ in the last step.

Recall the elementary fact
$(\sum_{i=1}^{m} y_{i})^{2}\leq m \sum_{i=1}^{m} y_{i}^{2}$,
which follows from the Cauchy-Schwarz inequality. In order to apply this
observation to the preceding expression, we estimate the number of terms in the inner sum.
The outer summation over $\setT$ can be restricted to non-empty sets $\setT$,
since $\bar{E}_{\phi}$ has no effect in the summation due to 
$\psi_{\setS}$ having zero mean. Thus, any given $\setT$ in the expression
has at least one element, and the sets $\setS\supset \setT$ in the 
collection $\collS$ must contain it; so the number of sets $\setS$ over
which the inner sum is taken cannot exceed $r$. Thus we have
\be\label{key-dec}\begin{split}
EU^{2} &\leq  \sum_{\setT} \,\,r
\sum_{\setS\supset \setT, \setS\in\collS} E\big(\Etb\psi_{\setS}(X_{\setS}) \big)^{2} \\
&= r\sumS \sum_{\setT\subset\setS}   
E\big( \Etb\psi_{\setS}(X_{\setS}) \big)^{2} \\
&= r\sum_{ \setS\in\collS} E\big(  \psi_{\setS}(X_{\setS}) \big)^{2} ,
\end{split}\ee
by rearranging the sums and using the orthogonality of the ANOVA decomposition again.
This proves the inequality.

Now suppose  $\psi_{\setS'}$ is
not additive. This means that for some set $\setT\subset\setS'$ with two elements,
$\Etb\psi_{\setS'}(X_{\setS'}) \neq 0$. Fix this choice of $\setT$.
Since $\collS$ is a discriminating collection,
not all of the at most ${r}$ subsets containing one element of $\setT$
can contain the other. Consequently, the inner sum in the inequality \eqref{key-dec}
runs over strictly fewer than ${r}$ subsets $\setS$, and the inequality \eqref{key-dec}
must be strict. Thus 
each $\psi_{\setS}$  must be an additive function if equality holds,
i.e., it must be composed only of main effects and no interactions.
\end{proof}

\par\vspace{.15in}
\noindent\begin{rmk}
The idea of the variance drop inequality goes back at least to Hoeffding's
seminal work \cite{Hoe48} on $U$-statistics.  
Suppose $\psi:\RL^{m}\ra\RL$ is symmetric
in its arguments, and $E\psi(X_{1},\ldots,X_{m})=0$.
Define
\be\label{U-defn}
U(X_{1},\ldots,X_{n})= \frac{1}{\binom{n}{m}} \sum_{\{\setS\subset[n]: |\setS|=m\}} \psi(X_{\setS}) .
\ee
Then Hoeffding \cite{Hoe48} showed
\be\label{Hoe-result}
EU^{2} \leq \frac{m}{n} E\psi^{2} ,
\ee
which is implied by Lemma~\ref{lem:vardrop} under the symmetry assumptions.
In statistical language, $U$ defined in \eqref{U-defn} is a $U$-statistic of 
degree $m$ with symmetric, mean zero kernel $\psi$ that
is applied to data of sample size $n$. Thus \eqref{Hoe-result}
quantitatively captures the reduction of variance of a $U$-statistic
when sample size $n$ increases. For $m=1$, this is the trivial fact that 
the empirical variance of a function based
on i.i.d. samples is the actual variance scaled by $n^{-1}$.
For $m>1$, the functions $\psi(X_{\setS})$ are no longer 
independent, nevertheless the variance of the $U$-statistic
drops by a factor of $\frac{m}{n}$. 
Our proof is valid for the more general non-symmetric case,
and also seems to illuminate the underlying statistical idea 
(the ANOVA decomposition) as well as the underlying geometry (Hilbert space projections) better 
than Hoeffding's original combinatorial proof. 
In \cite{ES81}, Efron and Stein 
assert in their Comment 3 that an ANOVA-like decomposition
``yields one-line proofs of Hoeffding's important theorems
5.1 and 5.2''; presumably our proof of Lemma~\ref{lem:vardrop}
is a generalization of what they had in mind. As mentioned before,
the application of such a variance drop lemma to information
inequalities was pioneered by ABBN \cite{ABBN04:1}.
They proved and used it in the case $\collS=\collS_{n-1}$
using clear notation that we adapt in developing our generalization 
above. A further generalization appears
when we consider refinements of our main inequalities
in Section~\ref{sec:ref}.
\end{rmk}
\par\vspace{.15in}

The third key tool in our approach to monotonicity is the well-known
link between Fisher information and entropy, whose origin is 
the de Bruijn identity first described by Stam \cite{Sta59}. 
This identity, which identifies the Fisher information as
the rate of change of the entropy on adding a normal, provides
a standard way of obtaining entropy inequalities from Fisher information
inequalities. 
An integral form of the de Bruijn identity was 
proved by Barron \cite{Bar86}. We express that integral 
in a form suitable for our purpose (cf., ABBN \cite{ABBN04:1} and
Verd\'u and Guo \cite{VG06}).

\par\vspace{.15in}
\noindent\begin{lem}\label{lem:deB}
Let $X$ be a random variable with a density and arbitrary finite variance.
Suppose $X_{t}\eqd X+\sqrt{t}Z$, where $Z$ is a standard normal
independent of $X$. Then,
\be\label{lemdeB}
H(X) = \half \log (2 \pi e)  - \half \int_0^{\infty} \bigg[I(X_t) - \frac{1}{1+t} \bigg] dt.
\ee
\end{lem}

\par\vspace{.15in}
\noindent\begin{proof}
In the case that the variances of $Z$ and $X$ match, equivalent forms of
this identity are given in \cite{Bar84:tr} and \cite{Bar86}.
Applying a change of variables using $t=\tau v$ to 
equation (2.23) of \cite{Bar84:tr} 
(which is also equivalent to equation (2.1)
of \cite{Bar86} by another change of variables), one has that
\ben
H(X) = \half \log (2 \pi e v) - \half \int_0^{\infty} \bigg[I(X_t) - \frac{1}{v+t}\bigg] dt,
\een
if $X$ has variance $v$ and $Z$ has variance 1. 
This has the advantage of positivity of the integrand but the 
disadvantage that it seems to depend on $v$. One can use 
\ben
\log v = \int_{0}^{\infty} \bigg[\frac{1}{1+t} -\frac{1}{v+t} \bigg] dt
\een
to re-express it in the form \eqref{lemdeB}, which does not depend on $v$.
\end{proof}

\section{Monotonicity in the IID case}
\label{sec:iid}

For clarity of presentation of ideas, we focus first on
the i.i.d. setting. For i.i.d. summands, inequalities \eqref{abbn-epi}
and \eqref{m-epi} reduce to the monotonicity property
$H(Y_{n})\geq H(Y_{m})$ for $n>m$,
where
\be\label{s-def}
Y_n =\frac{1}{\sqrt{n}} \Sumn X_i .
\ee 
We exhibit below how our approach provides a simple proof of this
monotonicity property, first proved by ABBN \cite{ABBN04:1}
using somewhat more elaborate means.
We begin by showing the monotonicity of the Fisher information.

\par\vspace{.15in}
\noindent\begin{prop}[{\sc Monotonicity of Fisher information}]\label{prop:I}
If $\{X_{i}\}$ are i.i.d. random variables, and $Y_{n-1}$ has an absolutely continuous density, then
\be\label{propI}
I(Y_{n})\leq I(Y_{n-1}),
\ee
with equality iff $X_{1}$ is normal or $I(Y_{n})=\infty$.
\end{prop}

\par\vspace{.15in}
\noindent\begin{proof}
We use the following notation:
The (unnormalized) sum is $S_{n}=\sum_{i\in[n]} X_{i}$,
and the leave-one-out sum leaving out $X_{j}$ is 
$S^{(j)}=\sum_{i\neq j} X_{i}$. 
Setting $\rho(S_{n})$ to be the score of $S_{n}$ and
$\rho_{j}$ to be the score of $S^{(j)}$, 
we have by Lemma~\ref{lem:conv} that 
$\rho(S_{n}) = E[ \rho_{j} | S_{n}]$
for each $j$, and hence
\ben
\rho(S_{n}) =  E \bigg[  \nth\sum_{j\in[n]}\rho_{j} \bigg| S_{n}  \bigg] .
\een
Since the norm of the score is not less than that of its projection 
(i.e., by the Cauchy-Schwarz inequality),
\ben
I(S_{n}) = E [\rho^{2}(S_{n})]
\leq  E \bigg[ \bigg( \nth\sum_{j\in[n]} \rho_{j} \bigg)^{2} \bigg].
\een
Lemma~\ref{lem:vardrop} yields
\ben 
E \bigg( \nth\sum_{j\in[n]} \rho_{j} \bigg)^{2} 
\leq (n-1) \sum_{j\in [n]} \frac{1}{n^{2}} E\rho_{j}^{2} 
= \frac{n-1}{n} I(S_{n-1}) ,
\een
so that
\ben
I(S_{n}) \leq \frac{n-1}{n} I(S_{n-1}) .
\een
If $X'=aX$, then $\rho_{X'}(X')=\frac{1}{a}\rho_{X}(X)$ and $a^{2}I(X')=I(X)$; hence
$$I(Y_{n})=nI(S_{n})\leq (n-1)I(S_{n-1})= I(Y_{n-1}). $$

The inequality implied by Lemma~\ref{lem:vardrop}
can be tight only if each $\rho_{j}$, considered as a function
of the random variables $X_{i}, i\neq j$, is additive.
However, we already know that $\rho_{j}$ is a function of the sum
of these random variables.
The only functions that are both additive and functions of the sum
are linear functions of the sum; hence the two sides of \eqref{propI}
can be finite and equal only if each of the scores $\rho_{j}$ is linear,
i.e., if all the $X_{i}$ are normal. It is trivial to check that $X_{1}$ normal or
$I(Y_{n})=\infty$ imply equality.
\end{proof}
\par\vspace{.15in}

The monotonicity result for entropy in the i.i.d. case now follows by combining
Proposition~\ref{prop:I} and Lemma~\ref{lem:deB}.

\par\vspace{.15in}
\noindent\begin{thm}[{\sc Monotonicity of Entropy: IID Case}]\label{thm:iid}
Suppose $\{X_{i}\}$ are i.i.d. random variables with densities
and finite variance. If the normalized sum $Y_{n}$ is defined by
\eqref{s-def}, then 
\ben
H(Y_{n})\geq H(Y_{n-1}) .
\een
The two sides are finite and equal iff $X_{1}$ is normal.
 \end{thm}
\par\vspace{.15in}

After the submission of these results to ISIT 2006 \cite{MB06:isit}, 
we became aware of a contemporaneous and independent development 
of the simple proof of the monotonicity fact (Theorem~\ref{thm:iid}) 
by Tulino and Verd\'u \cite{TV06}.  In their work they take 
nice advantage of projection properties through minimum mean squared error 
interpretations. It is pertinent to note that the proofs of Theorem 1 
(in \cite{TV06} and in this paper) share essentials, because 
of the following observations.

Consider estimation of a random variable $X$ from an observation $Y=X+Z$ 
in which an independent standard normal $Z$ has been added.  
Then the score function of Y is related to the difference between two
predictors of X (maximum likelihood and Bayes), i.e.,
\be
- \rho(Y) = Y - E[X|Y] ,
\ee
and hence the Fisher information $I(Y)=E \rho^2 (Y)$
is the same as the mean square difference
$E[(Y-E[X|Y])^2]$, or equivalently, by the Pythagorean identity,
\be
I(Y)= \VAR(Z) - E[(X-E[X|Y])^2].
\ee
Thus the Fisher information (entropy derivative) is related to the
minimal mean squared error.  These (and more general) identities
relating differences between predictors to scores and relating their
mean squared errors to Fisher informations are developed in statistical
decision theory in the work of Stein and Brown. These developments
are described, for instance, in the point estimation text by 
Lehmann and Casella \cite{LC98:book}[Chapters 4.3 and 5.5], 
in their study of Bayes risk, admissibility and
minimaxity of conditional means $E[X|Y]$.

Tulino and Verd\'u \cite{TV06} emphasize the minimal mean squared 
error property of the entropy derivative and associated projection properies that
(along with the variance drop inequality which they note in the leave-one-out 
case) also give Proposition~\ref{prop:I} and Theorem~\ref{thm:iid}.  
That is a nice idea. Working directly with the minimal mean squared error as the
entropy derivative they bypass the use of Fisher information.  In the
same manner Verd\'u and Guo \cite{VG06} give an alternative proof 
of the Shannon-Stam entropy power inequality.  If one takes note 
of the above identities one sees
that their proofs and ours are substantially the same, except that the
same quantities are given alternative interpretations in the two works,
and that we give extensions to arbitrary collections of subsets.

\section{Fisher Information Inequalities}
\label{sec:fii}

In this section, we demonstrate our core inequality \eqref{central-I}.

\par\vspace{.15in}
\noindent\begin{prop}\label{prop:inid-I}
Let $\{X_{i}\}$ be independent random variables with densities
and finite variances. Define
\be\label{def:T}
T_{n} = \sum_{i\in[n]} X_{i} \hspace{0.4in}\text{ and }\hspace{0.4in} T^{(\setS)}= \sum_{i\in \setS} X_{i} ,
\ee
for each $\setS\in\collS$, where $\collS$ is an arbitrary collection of subsets of $[n]$.
Let $w$ be any probability distribution on $\collS$.
If each $T^{(\setS)}$ has an absolute continuous density, then
\be
I(T_{n})\leq {r} \sumS \ws^{2} I(T^{(\setS)}) ,
\ee
where $\ws=w(\{\setS\})$. When $\collS$ is discriminating, 
both sides can be finite and equal only if each $X_{i}$ is normal.
\end{prop}

\par\vspace{.15in}
\noindent\begin{proof}
Let $\rho_{\setS}$ be the score of $T^{(\setS)}$.
We proceed in accordance with the outline in the introduction.
Indeed, Lemma~\ref{lem:conv} implies that
$\rho(T_{n}) =  E [\rho_{\setS} | T_{n}]$
for each $\setS$. Taking a convex combinations of these 
identities gives, for any $\{\ws\}$
such that $\sum_{\setS\in\collS } \ws=1$,
\be\begin{split}
\rho(T_{n}) = \sum_{\setS\in\collS } \ws E [\rho_{\setS} | T_{n}]  
= E \bigg[  \sum_{\setS\in\collS } \ws \rho_{\setS} \bigg| T_{n}  \bigg] .
\end{split}\ee
Applying the Cauchy-Schwarz inequality,
\be
\rho^{2}(T_{n}) \leq E \bigg[  \bigg(\sum_{\setS\in\collS } \ws \rho_{\setS}\bigg)^{2} \bigg| T_{n}  \bigg] .
\ee
Taking the expectation and then applying Lemma~\ref{lem:vardrop} 
in succession, we get
\bea
\label{bd3}
E\big[\rho^{2}(T_{n})\big]  &\leq E \bigg( \sum_{\setS\in\collS } \ws \rho_{\setS} \bigg)^{2} \\
\label{bd4}
&\leq {r} \sum_{\setS\in\collS } E(\ws \rho_{\setS})^{2} \\
&= {r} \sumS \ws^{2} I(T^{(\setS)}) ,
\ee
as desired. The application of Lemma~\ref{lem:vardrop} can yield
equality only if each $\rho(T^{(\setS)})$ is additive;
since the score $\rho(T^{(\setS)})$ is already a function of the sum $T^{(\setS)}$,
it must in fact be a linear function, so that each $X_{i}$ must be normal.
\end{proof}
\par\vspace{.15in}

Naturally, it is of interest to minimize the upper bound of Proposition~\ref{prop:inid-I} 
over the weighting distribution $w$, which is easily done either
by an application of Jensen's inequality for the reciprocal function,
or by the method of Lagrange multipliers. Optimization of the bound
implies that Proposition~\ref{prop:inid-I} is equivalent to the 
following Fisher information inequalities. 

\par\vspace{.15in}
\noindent\begin{thm}\label{thm:FII}
Let $\{X_{i}\}$ be independent random variables such that
each $T^{(\setS)}$ has an absolutely continuous density. 
Then
\be\label{fii}
\frac{1}{I(T_{n})} \geq  \frac{1}{{r}} \sumS \frac{1}{I(T^{(\setS)})} .
\ee
When $\collS$ is discriminating, the two sides are 
positive and equal iff each $X_{i}$ is normal
and $\collS$ is also balanced.
\end{thm}
\par\vspace{.15in}

\begin{rmk}
Theorem~\ref{thm:FII} for the special case $\collS=\collS_{1}$ of singleton
sets is sometimes known as 
the ``Stam inequality'' and has a long history. Stam \cite{Sta59}
was the first to prove Proposition~\ref{prop:inid-I} for $\collS_{1}$, and he credited
his doctoral advisor de Bruijn with noticing the equivalence to 
Theorem~\ref{thm:FII} for $\collS_{1}$. Subsequently 
several different proofs have appeared:
in Blachman \cite{Bla65} using Lemma~\ref{lem:conv}, 
in Carlen \cite{Car91} using another superadditivity property of the Fisher information,  
and in Kagan \cite{Kag02} as a consequence of 
an inequality for Pitman estimators. On the other hand, the special case of 
the leave-one-out sets $\collS=\collS_{n-1}$ in
Theorem~\ref{thm:FII} was first proved in ABBN \cite{ABBN04:1}. 
Zamir \cite{Zam98} used data processing 
properties of the Fisher information to prove some different extensions 
of the $\collS_{1}$ case, including
a multivariate version; see also Liu and Viswanath \cite{LV05:pre}
for some related interpretations.
Our result for arbitrary collections of subsets is new; yet our proof
of this general result is essentially no harder than the elementary 
proofs of the original inequality by Stam and Blachman.
\end{rmk}
\par\vspace{.15in}
\begin{rmk}\label{rmk:cltgap}
While inequality \eqref{bd3} in the proof above uses a Pythagorean inequality,
one may use the associated Pythagorean identity to characterize the difference
as the mean square of $\sum_{\setS} \ws\rho_{\setS}-\rho(T_{n})$.
In the i.i.d. case with $n=2m$ and $\collS$ a disjoint pair of subsets
of size $m$, this drop in Fisher distance from the normal played an 
essential role in the previously mentioned CLT analyses of \cite{Shi75, Bro82, Bar86, JB04}. 
Furthermore for general $\collS$, we have from the variance drop analysis that the 
gap in inequality \eqref{bd4} is characterized by the non-additive ANOVA
components of the score functions. We point out these observations as an 
encouragement to examination of the information drop for collections such
as $\collS_{n-1}$ and $\collS_{n/2}$ in refined analysis of CLT rates
under more general conditions.
\end{rmk}

\section{Entropy Inequalities}
\label{sec:ent}

The Fisher information inequality of the previous section
yields a corresponding entropy inequality. 

\par\vspace{.15in}
\noindent\begin{prop}[{\sc Entropy of Sums}]\label{prop:sums}
Let $\{X_{i}\}$ be independent random variables with densities. Then,
for any probability distribution $w$ on $\collS$ such that $\ws\leq \rth$
for each $\setS$,
\be\label{H-sums}\begin{split}
H\bigg(\sum_{i\in[n]} X_{i}\bigg) \geq \sum_{\setS\in\collS } 
& \ws H\bigg(\sum_{i\in\setS} X_{i}\bigg) \\
& + \half H(w) -\half \log  {r} ,
\end{split}\ee
where $H(w)=\sumS \ws\log\frac{1}{\ws}$
is the discrete entropy of $w$.
When $\collS$ is discriminating, equality can hold
only if each $X_{i}$ is normal.
\end{prop}
\par\vspace{.15in}

\noindent\begin{proof}
As pointed out in the Introduction, Proposition~\ref{prop:inid-I} is equivalent to
\ben\begin{split}
I\bigg(\sum_{i\in[n]} Y_{i}\bigg) 
&\leq \sumS \ws I\bigg( \frac{\sum_{i\in\setS} Y_{i}}{\sqrt{\ws {r}}} \bigg) ,
\end{split}\een
for independent random variables $Y_{i}$.
For application of the Fisher information inequality to 
entropy inequalities we also need for an independent 
standard normal $Z$ that 
\be\label{pert-I}
I(T_n + \sqrt{\tau}Z) \leq \sumS \ws I(\frac{T^{\setS}}{\sqrt{r\ws}}+ \sqrt{\tau} Z), 
\ee
at least for suitable values of $\ws$. We will show below that this 
holds when $r\ws\leq 1$ for each $\setS$, and thus
\ben\begin{split}
H(T_{n}) &= \half\log (2\pi e) - \half\int_{0}^{\infty}
\bigg( I(T_{[n]}+\sqrt{\tau}Z)-\frac{1}{1+\tau} \bigg) dt \\
&\geq \sumS\ws \bigg[ \half\log (2\pi e) - \\
&\quad\quad\quad \half\int_{0}^{\infty}
\bigg\{ I\bigg( \frac{T^{\setS}}{\sqrt{r\ws}} + \sqrt{\tau} Z\bigg)-\frac{1}{1+\tau} \bigg\} dt \bigg] \\
&=  \sumS \ws H\bigg( \frac{T^{\setS}}{\sqrt{r\ws}} \bigg) ,
\end{split}\een
using \eqref{pert-I} for the inequality and Lemma 3 for the equalities.
By the scaling property of entropy, this implies
\ben
H(T_{n}) 
\geq  \sum_{\setS\in\collS } \ws H(T^{\setS})
- \half \sum_{\setS\in\collS } \ws\log\ws    -\half\log {r}  ,
\een
proving the desired result.

The inequality \eqref{pert-I} is true though 
not immediately so (the naive approach of adding an 
independent normal to each $X_i$ does not work to get 
our desired inequalities when the subsets have more than 
one element).  What we need is to provide a collection of 
independent normal random variables $Z_j$ for some set of 
indices $j$ (possibly many more than $n$ of them). For each 
$\setS$ in $\collS$ we need an assignment of subset sums of $Z_j$ (called 
say $Z^{\setS'}$) which has variance $r\ws$, such that no $j$
is in more than $r$ of the subsets $\setS'$. Then by 
Proposition~\ref{prop:inid-I} (applied to the collection $\collS'$ of augmented sets 
$\setS\cup\setS'$ for each $\setS$ in $\collS$) we have 
\ben
I(T_n + \sqrt{t} Z) \leq \sumS r \ws^2 I(T^{\setS} + \sqrt{t} Z^{\setS'}) 
\een
from which the desired inequality follows using the fact 
that $a^2 I(X) = I(X/a)$.  
Assuming that $r\ws \leq 1$ (which 
will be sufficient for our needs), we provide such a 
construction of $Z_j$ and their subset sums in the case of rational 
weights, say $\ws = W(\setS)/M$, where the denominator $M$ may be 
large.  Indeed set $Z_1, \ldots, Z_M$ independent mean-zero normals 
each of variance $1/M$.  For each $\setS$ we construct a set $\setS'$ that has 
precisely $rW(\setS)$ normals and each normal is assigned to 
precisely $r$ of these sets.  This may be done systematically by 
considering the sets $\setS_{1}, \setS_{2}, \ldots$ in $\collS$ in some order.  We 
let $\setS'_{1}$ be the the first $rW(\setS_1)$ indices $j$ (not more 
than $M$ by assumption), we let $\setS'_{2}$ be the next $rW(\setS_2)$ indices 
(looping back to the first index once we pass $M$) and so on.
This proves the validity of \eqref{pert-I} for rational weights;
its validity for general weights follows by continuity.
\end{proof}

\par\vspace{.15in}
\noindent\begin{rmk}
One may re-express the inequality of Proposition~\ref{prop:sums}
as a statement for the relative entropies with
respect to normals of the same variance. 
If $X$ has density $f$, we write 
$D(X)=E[\log \frac{f(X)}{g(X)}]= H(Z)-H(X)$,
where $Z$ has the Gaussian density $g$ with the same variance as $X$. 
Then,
for any probability distribution $w$ on a balanced collection $\collS$,
\be\label{relent}\begin{split}
D\bigg(\sum_{i\in[n]} X_{i}\bigg) \leq \sum_{\setS\in\collS } 
& \ws D\bigg(\sum_{i\in\setS} X_{i}\bigg) 
+ \half D(w\|\eta) ,
\end{split}\ee
where $\eta$ is the probability distribution on $\collS$ given by
$\eta_{\setS}=\frac{\text{Var}(T^{(\setS)})}{r\text{Var}(T_{n})}$,
and $D(w\|\eta)$ is the (discrete) relative entropy of $w$ with respect to $\eta$.
When $\collS$ is also discriminating, equality holds
iff each $X_{i}$ is normal.
Theorem 1 of Tulino and Verd\'u \cite{TV06}
is the special case of inequality \eqref{relent} for the collection of leave-one-out sets.
Inequality \eqref{relent} can be further extended to the case 
where $\collS$ is not balanced, but in that case $\eta$ is a subprobability distribution.
The conclusions \eqref{H-sums} and \eqref{relent} are equivalent, and as seen in the next 
section, are equivalent to our subset sum entropy power inequality.
\end{rmk}
\par\vspace{.15in}
\begin{rmk}
It will become evident in the next section that
the condition $r\ws\leq 1$ in Proposition~\ref{prop:sums}
is not needed for the validity of the conclusions \eqref{H-sums} and 
\eqref{relent} (see Remark~\ref{rmk:noass}). 
\end{rmk}

\section{Entropy power inequalities}
\label{sec:epi}

Proposition~\ref{prop:sums} is equivalent to a subset sum
entropy power inequality. Recall that the entropy power $\frac{e^{2H(X)}}{2\pi e}$
is the variance of the normal with the same entropy as $X$. 
The term entropy power is also used for the quantity 
\be
\calN(X)=e^{2H(X)} ,
\ee
even when the constant factor of $2\pi e$ is excluded.

\par\vspace{.15in}
\noindent\begin{thm}\label{thm:epi}
For independent random variables with finite variances, 
\be\label{epi-std}
\calN\bigg(\sum_{i\in [n]} X_{i}\bigg)  \geq \frac{1}{{r}} 
\sum_{\setS\in\collS } \calN\bigg(\sum_{i\in\setS} X_{i}\bigg) .
\ee
When $\collS$ is discriminating, the two sides are 
equal iff each $X_{i}$ is normal
and $\collS$ is also balanced.
\end{thm}
\par\vspace{.15in}

\noindent\begin{proof}
Let $T^{\setS}$ be the subset sums as defined in \eqref{def:T}.
Define
${\bf Z}=\sum_{\setS\in\collS } \calN(T^{\setS})$
as the normalizing constant producing the weights
\be\label{mu-defn}
\mus=\frac{\calN(T^{\setS})}{{\bf Z}} . 
\ee

If $N(T^{\setS}) > {\bf Z}/r$ for some $\setS\in\collS$ then we trivially 
have $N(T_n) \geq N(T^{\setS}) > {\bf Z}/r$ which is the desired result, so 
now assume $N(T^{\setS}) \leq {\bf Z}/r$ for all $\setS\in\collS$, that 
is, $r \mu_{\setS} \leq 1$ for each $\setS$.

Since 
\ben
H(T^{(\setS)})=\half \log \calN(T^{\setS})=\half\log [\mus {\bf Z}] ,
\een
Proposition~\ref{prop:sums} implies for any weighting distribution
$w$ with $r\ws\leq 1$ that
\be\label{H-case2}\begin{split}
H(T_{n})&\geq \half \sum_{\setS\in\collS } \ws \log [\mus {\bf Z}]+ \half \bigg[ H(w) - \log  {r} \bigg] \\
&= \half \bigg[\log \frac{{\bf Z}}{{r}} -  D(w\|\mu) \bigg],
\end{split}\ee
where $D(w\|\mu)$ is the (discrete) relative entropy.
Exponentiating gives
\be\label{epi-1}
\calN(T_{n}) \geq { {r}}^{-1} e^{-D(w\|\mu)} {\bf Z} .
\ee
It remains to optimize the right side over $w$, or equivalently to
minimize $D(w\|\mu)$ over feasible $w$. Since 
$r\mus\leq 1$ by assumption, $w=\mu$ is a feasible choice, yielding
the desired inequality. 

The necessary conditions for equality follow from that for 
Proposition~\ref{prop:sums}, and it is easily checked using 
Fact~\ref{f:add} that this is also sufficient.
The proof is complete.
\end{proof}

\par\vspace{.15in}
\noindent\begin{rmk}
To understand this result fully, it is useful to note that if a discriminating collection
$\collS$ is not balanced, it can always be augmented to a new collection
$\collS'$ that is balanced in such a way that the inequality \eqref{epi-std}
for $\collS'$ becomes strictly better than that for $\collS$. Indeed, 
if index $i$ appears in $r(i)$ sets of $\collS$, one can always find $r-r(i)$
sets of $\collS$ not containing $i$ (since $r\leq |\collS|$), and add $i$
to each of these sets. The inequality \eqref{epi-std} for $\collS'$
is strictly better since this collection has the same $r$ and the subset sum
entropy powers on the right side are higher due to the addition of independent
random variables. While equality in \eqref{epi-std} is impossible for the 
unbalanced collection $\collS$, it holds for normals for the augmented, balanced
collection $\collS'$. This illuminates the conditions for equality in Theorem~\ref{thm:epi}.
\end{rmk}

\par\vspace{.15in}
\noindent\begin{rmk}
The traditional Shannon inequality involving the entropy powers of the
summands \cite{Sha48} as well as the inequality of ABBN \cite{ABBN04:1}
involving the  entropy powers of the ``leave-one-out'' 
normalized sums are two special cases of Theorem~\ref{thm:epi},
corresponding to $\collS=\collS_{1}$ and $\collS=\collS_{n-1}$. 
Proofs of the former subsequent to Shannon's include those of Stam \cite{Sta59},
Blachman \cite{Bla65}, Lieb \cite{Lie78} (using Young's inequality
for convolutions with the sharp constant), 
Dembo, Cover and Thomas \cite{DCT91} (building on
Costa and Cover \cite{CC84}),
and Verd\'u and Guo \cite{VG06}.
Note that unlike the previous proofs of these special cases,
our proof of the equivalence between the linear form of Proposition~\ref{prop:sums} 
and Theorem~\ref{thm:epi} reduces to the 
non-negativity of the relative entropy.
\end{rmk}

\par\vspace{.15in}
\noindent\begin{rmk}\label{rmk:noass}
To see that \eqref{H-sums} (and hence  \eqref{relent})
holds without any assumption on $w$, simply note that when
the assumption is not satisfied, the entropy power inequality of 
Theorem~\ref{thm:epi} implies trivially that
\ben
\calN(T_{n})\geq r^{-1}{\bf Z}\geq 
r^{-1} e^{-D(w\|\mu)} {\bf Z} ,
\een
for $\mu$ defined by \eqref{mu-defn},
and inverting the steps of \eqref{H-case2}
yields \eqref{H-sums}. 
\end{rmk}

\section{Entropy is Monotone on Average}
\label{sec:avg}

In this section, we consider the behavior of the entropy
of sums of independent but not necessarily identically distributed 
(i.n.i.d.) random variables under various scalings.

First we look at sums scaled according to the number of summands. 
Fix the collection $\collS_{m}=\{\setS\subset[n]:|\setS|=m\}$.
For i.n.i.d. random variables $X_{i}$, let
\be\label{def:Y}
Y_{n}=\frac{\sum_{i\in[n]} X_{i}}{\sqrt{n}}
\hspace{0.3in}\text{ and }\hspace{0.3in} 
Y_{m}^{(\setS)}=\frac{\sum_{i\in\setS} X_{i}}{\sqrt{m}}
\ee
for $\setS\in\collS_{m}$ be the scaled sums. 
Then Proposition~\ref{prop:sums} applied to $\collS_{m}$ implies
\ben
H(Y_{n}) \geq \sum_{\setS\in\collS_{m}} \ws H(Y_{m}^{(\setS)})
- \half \bigg[\log \binom{n}{m} -  H(w) \bigg] .
\een
The term on the right indicates that we pay a cost for deviations of the
weighting distribution $w$ from the uniform.
In particular, choosing $w$ to be uniform 
implies that entropy is ``monotone on average'' with uniform weights
for scaled sums of i.n.i.d. random variables. Applying Theorem~\ref{thm:epi}
to $\collS_{m}$ yields a similar conclusion for entropy power. These
observations, which can also be deduced from the results of ABBN \cite{ABBN04:1}, 
are collected in Corollary~\ref{cor:avgmon}.

\par\vspace{.15in}
\noindent\begin{cor}\label{cor:avgmon}
Suppose $X_{i}$ are independent random variables with densities, 
and the scaled sums are defined by \eqref{def:Y}.
Then 
\be\begin{split}
H(Y_{n}) &\geq \frac{1}{\binom{n}{m}} \sum_{\setS\in\collS_{m}} H(Y_{m}^{(\setS)}) \\
\text{and } \calN(Y_{n}) &\geq \frac{1}{\binom{n}{m}} \sum_{\setS\in\collS_{m}} \calN(Y_{m}^{(\setS)}) .
\end{split}\ee
\end{cor}

\par\vspace{.15in}
\noindent\begin{rmk}
It is interesting to contrast Corollary~\ref{cor:avgmon}
with the following results of Han \cite{Han78} and Dembo, Cover and Thomas \cite{DCT91}.
With no assumptions on $(X_{1},\ldots,X_{n})$ except that they
have a joint density, the above authors show that
\be
\frac{H(X_{[n]})}{n} \leq \frac{1}{\binom{n}{m}} \sum_{\setS\in\collS_{m}} \frac{H(X_{\setS})}{m}
\ee
and
\be
\big[ \calN (X_{[n]}) \big]^{\frac{1}{n}} \leq  \frac{1}{\binom{n}{m}} 
\sum_{\setS\in\collS_{m}} \big[ \calN (X_{\setS}) \big]^{\frac{1}{m}} .
\ee
where $H(X_{\setS})$ and $\calN(X_{\setS})$ denote the joint entropy and 
joint entropy power of $X_{\setS}$ respectively. These bounds have a form 
very similar to that of Corollary~\ref{cor:avgmon}.
In fact, such an analogy between inequalities for entropy of sums and
joint entropies goes much deeper (so that all of the entropy power
inequalities we present here have analogues for joint entropy).
More details can be found in Madiman and Tetali \cite{MT07:isit}.
\end{rmk}
\par\vspace{.15in}


Next, we consider sums of independent random variables standardized by
their variances. This is motivated by the following consideration.
Consider a sequence of i.n.i.d. random variables $\{X_{i}:i\in\Nat\}$
with zero mean and finite variances, $\sigma_{i}^{2}=\VAR(X_{i})$.
The variance of the sum of $n$ variables is denoted 
$v_{n}=\sum_{i\in[n]} \sigma_{i}^{2}$, and the
standardized sum is
\ben
V_{n}= \frac{\sum_{i\in[n]} X_{i}}{\sqrt{v_{n}}} .
\een
The Lindeberg-Feller central limit theorem gives conditions under which
$V_{n}\Ra N(0,1)$. Johnson \cite{Joh00} has proved an entropic version
of this classical theorem, showing (under appropriate conditions)
that $H(V_{n})\ra \half\log(2\pi e)$ and hence the relative entropy from the unit
normal tends to 0. Is there an analogue of the monotonicity of information in 
this setting?

We address this question in the following theorem, and give two proofs.
The first proof is based on considering appropriately standardized linear
combinations of independent random variables, 
and generalizes Theorem 2 of ABBN \cite{ABBN04:1}. The second
proof is outlined in Remark~\ref{rmk:altpf}.

\par\vspace{.15in}
\noindent\begin{thm}[{\sc Monotonicity on Average}]\label{thm:avg}
Suppose $\{X_{i}:i\in[n]\}$ are  independent random variables
with densities, and $X_{i}$ has finite variance 
$\sigma_{i}^{2}$. Set $v_{n}=\sum_{i\in[n]} \sigma_{i}^{2}$
and $v_{\setS}=\sum_{i\in\setS} \sigma_{i}^{2}$ for sets $\setS$
in the balanced collection $\collS$. Define the standardized sums 
\be
V_{n}= \frac{\sum_{i\in[n]} X_{i}}{\sqrt{v_{n}}} 
\ee
and
\be
V^{(\setS)}= \frac{\sum_{i\in\setS} X_{i}}{\sqrt{v_{\setS}}} .
\ee
Then
\be
H(V_{n})\geq \sumS \eta_{\setS} H(V^{(\setS)}) ,
\ee
where $\eta_{\setS}= \frac{1}{{r}} \frac{v_{\setS}}{v_{n}}$.
Furthermore, if $\collS$ is also discriminating, then the inequality is strict
unless each $X_{i}$ is normal.  
\end{thm}

\par\vspace{.15in}
\noindent\begin{proof}
Let $a_{i}, i\in [n]$ be a collection of non-negative real numbers such that 
$ \sum_{i=1}^{n} a_{i}^{2}=1$. Define 
$\asb= \big[\sum_{i\in \setS} a_{i}^{2}\big]^{\half}$
and the weights $\lam_{\setS}= \frac{\asb^{2}}{{r}}$ for $\setS\in\collS$.
Applying the inequalities of Theorem~\ref{thm:FII},
Proposition~\ref{prop:sums} and Theorem~\ref{thm:epi}
to independent random variables $a_{i}X_{i}'$, and utilizing the scaling 
properties of the relevant information quantities, one finds that
\be\label{linc}
\psi\bigg(\sum_{i=1}^{n} a_{i} X_{i}'\bigg)\geq 
\sumS \lam_{\setS} \psi\bigg( \frac{1}{\asb} \sum_{i\in \setS} a_{i} X_{i}' \bigg) ,
\ee
where $\psi$ represents either the inverse Fisher information $I^{-1}$
or the entropy $H$ or the entropy power $\calN$.

The conclusion of Theorem~\ref{thm:avg} is a particular instance of \eqref{linc}.
Indeed, we can express the random variables of interest as $X_{i}=\sigma_{i}X_{i}'$, 
so that each $X_{i}'$ has variance 1. Choose $a_{i}=\frac{\sigma_{i}}{\sqrt{v_{n}}}$, 
which is valid since $\sum_{i\in[n]} a_{i}^{2}=1$.
Then $\asb^{2}=\sum_{i\in\setS} a_{i}^{2}=\frac{v_{\setS}}{v_{n}}$
and $\lam_{\setS}=\eta_{\setS}$. Thus
\ben
V_{n}= \sum_{i\in[n]} \frac{X_{i}}{\sqrt{v_{n}}} = \sum_{i\in[n]} a_{i} X_{i}' ,
\een
and
\ben
V^{(\setS)}= \sum_{i\in\setS} \frac{ X_{i}}{\sqrt{v_{\setS}}} 
= \frac{1}{\asb}\sum_{i\in\setS} a_{i}X_{i}' .
\een
Now an application of \eqref{linc} gives the desired result, not just for $H$
but also for $\calN$ and $I^{-1}$.
\end{proof}

\par\vspace{.15in}
\noindent\begin{rmk}
Since the collection $\collS$ is balanced,
it follows from Fact~\ref{f:add} that
$\eta_{\setS}$ defines a probability distribution on $\collS$.
This justifies the interpretation of Theorem~\ref{thm:avg} as 
displaying ``monotonicity on average''. The averaging distribution
$\eta$ is tuned to the random variables of interest, through their variances.
\end{rmk}
\par\vspace{.15in}

\noindent\begin{rmk}\label{rmk:altpf}
Theorem~\ref{thm:avg} also follows directly from \eqref{relent}
upon setting $\ws=\eta_{\setS}$ and noting that the definition of $D(X)$
is scale invariant (i.e., $D(aX)=D(X)$ for any real number $a$).
\end{rmk}
\par\vspace{.15in}

Let us briefly comment on the interpretation of this result. 
As discussed before, when the summands are i.i.d., 
entropic convergence of $V_{n}$ to the normal was shown in \cite{Bar86}, 
and ABBN \cite{ABBN04:1} showed that this sequence of entropies
is monotonically increasing. This completes a long-conjectured intuitive
picture of the central limit theorem: 
forming normalized sums that keep the variance constant yields random variables
with increasing entropy, and this sequence of entropies converges to the maximum
entropy possible, which is the entropy of the normal with that variance. In this
sense, the CLT is a formulation of the ``second law of thermodynamics'' in physics.
Theorem~\ref{thm:avg} above shows that even in the setting of variance-standardized sums
of i.n.i.d. random variables,
a general monotonicity on average property holds with respect to 
an arbitrary collection of normalized subset sums.
This strengthens the  ``second law'' interpretation of central limit theorems.

A similar monotonicity on average property also holds for appropriate
notions of Fisher information in convergence of sums of discrete random 
variables to the Poisson and compound Poisson distributions; 
details may be found in \cite{MJK07:isit}.

\section{A Refined Inequality}
\label{sec:ref}

Various extensions of the basic inequalities presented above are possible;
we present one here. To state it, we find it
convenient to recall the notion of a fractional packing from 
discrete mathematics (see, e.g., 
Chung, F\"uredi, Garey and Graham \cite{CFGG88}).

\par\vspace{.15in}
\begin{defn}
Let $\collS$ be a collection of subsets of $[n]$. 
A collection $\{\bs:\setS\in\collS\}$ of non-negative real numbers
is called a fractional packing for $\collS$ if
\be\label{def:FP}
\sum_{\setS\ni i, \setS\in\collS} \bs \leq 1 
\ee 
for each $i$ in $[n]$.
\end{defn}
\par\vspace{.15in}

Note that if the numbers $\bs$ are constrained to only take the values
0 and 1, then the condition above entails that not more than one set
in $\collS$ can contain $i$, i.e., that the sets $\setS\in\collS$
are pairwise disjoint, and provide a packing of the set $[n]$. 
We may interpret a fractional packing as a ``packing'' of $[n]$ using sets
in $\collS$, each of which contains only a fractional piece (namely, $\beta_{\setS}$)
of the elements in that set. 

We now present a refined version of Lemma~\ref{lem:vardrop}.

\par\vspace{.15in}
\noindent\begin{lem}[{\sc Variance drop: General version}]\label{lem:vardrop-gen}
Suppose $U(X_{1},\ldots,X_{n})$ is a $\collS$-additive function
with mean zero components, as in Lemma~\ref{lem:vardrop}.
Then
\be
EU^{2} \leq \sumS 
\frac{1}{\bs} E \big( \psi_{\setS}(X_{\setS}) \big)^{2} ,
\ee
for any fractional packing $\{\bs:\setS\in\collS\}$.
\end{lem}
\par\vspace{.15in}

\begin{proof}
As in the proof of  Lemma~\ref{lem:vardrop}, we have
\ben\begin{split}
EU^{2} &=  \sum_{\setT} E\bigg[ 
\sum_{\setS\supset \setT, \setS\in\collS}  \Etb\psi_{\setS}(X_{\setS}) \bigg]^{2} .
\end{split}\een
We now proceed to perform a different estimation of this expression,
recalling as in the proof of Lemma~\ref{lem:vardrop}, that
the outer summation over $\setT$ can be restricted to non-empty sets $\setT$.
By the Cauchy-Schwarz inequality,
\ben\begin{split}
\bigg[&\sum_{\setS\supset \setT, \setS\in\collS} \Etb\psi_{\setS}(X_{\setS}) \bigg]^{2} \leq \\
&\quad \bigg[\sum_{\setS\supset \setT, \setS\in\collS} (\sqrt{\bs})^{2}\bigg] 
\bigg[\sum_{\setS\supset \setT, \setS\in\collS} \bigg( 
\frac{1}{\sqrt{\bs}} \Etb\psi_{\setS}(X_{\setS}) \bigg)^{2}\bigg] .
\end{split}\een
Since any $\setT$ of interest has at least one element, the 
definition of a fractional packing implies that
\ben
\sum_{\setS\supset \setT, \setS\in\collS} \bs \leq 1.
\een
Thus
\be\begin{split}
EU^{2}&\leq \sum_{\setT} \sum_{\setS\supset \setT, \setS\in\collS} 
\frac{1}{\bs} E\big( \Etb\psi_{\setS}(X_{\setS}) \big)^{2} \\
&=  \sumS \frac{1}{\bs} 
\sum_{\setT\subset \setS} E\big( \Etb\psi_{\setS}(X_{\setS}) \big)^{2} \\
&=  \sumS \frac{1}{\bs} E\big(\psi_{\setS}(X_{\setS}) \big)^{2} .
\end{split}\ee
\end{proof}

Exactly as before, one can obtain inequalities for Fisher information, entropy and
entropy power based on this form of the variance drop lemma. The idea of looking 
for such refinements with coefficients depending on $\setS$
arose in conversations with Tom Cover and Prasad Tetali at ISIT 2006 in Seattle,
after our basic results described in the previous sections were presented. 
In particular, Prasad's joint work with one of us \cite{MT07:isit} influenced the development of 
Theorem~\ref{thm:FII-gen}.  

\par\vspace{.15in}
\begin{thm}\label{thm:FII-gen}
Let $\{\bs:\setS\in\collS\}$ be any fractional packing for $\collS$.
Then 
\be\label{our-epi-ref}
I^{-1}(X_{1}+\ldots+X_{n}) \geq \sumS 
\bs I^{-1} \big( \sum_{j\in\setS} X_{j}\big) .
\ee
\end{thm}
\par\vspace{.15in}

For given subset sum informations, the best such lower bound
on the information of the total sum would involve maximizing the right side
of \eqref{our-epi-ref} subject to the linear constraints \eqref{def:FP}.
This linear programming problem, a version of which is the problem of optimal fractional packing
well-studied in combinatorics (see, e.g., \cite{CFGG88}), does not have an explicit solution
in general.

A natural choice of a fractional packing in Theorem~\ref{thm:FII-gen} 
leads to the following corollary. 

\par\vspace{.15in}
\begin{cor}\label{cor:rs}
For any collection $\collS$ of subsets of $[n]$, let $r(i)$
denote the number of sets in $\collS$ that contain $i$.
In the same setting as Theorem~\ref{thm:FII}, we have
\be\label{fii-rs}
\frac{1}{I(X_{1}+\ldots+X_{n})} \geq \sumS 
\frac{1}{r(\setS) I\big( \sum_{j\in\setS} X_{j}\big)} ,
\ee
where $r(\setS)$ is the maximum value of $r(i)$ over the indices
$i$ in $\setS$.  

We say that $\collS$ is quasibalanced
if $r(i)=r(\setS)$ for each $i\in\setS$ and each $\setS$.
If $\collS$ is discriminating, equality holds in \eqref{fii-rs}
if the $X_{i}$ are normal and $\collS$ is quasibalanced.
\end{cor}
\par\vspace{.15in}

\par\vspace{.15in}
\begin{rmk}
For any collection $\collS$ and any $\setS\in\collS$, 
$r(\setS)\leq r$ by definition. Thus Theorem~\ref{thm:FII-gen}
and Corollary~\ref{cor:rs} generalize Theorem~\ref{thm:FII}.
Furthermore, from the equality conditions in Corollary~\ref{cor:rs},
we see that equality can hold in these more general inequalities even
for collections $\collS$ that are not balanced, which was not possible with the
original formulation in Theorem~\ref{thm:FII}.
\end{rmk}

\par\vspace{.15in}
\begin{rmk}
One can also give an alternate proof of Theorem~\ref{thm:FII-gen} 
using Corollary~\ref{cor:rs} (which could be proved directly), 
so that the two results are mathematically 
equivalent. The key to doing this is the observation that nowhere in our
proofs do we actually require that the sets $\setS$ in $\collS$ are distinct.
In other words, given a collection $\collS$, one may look at an augmented 
collection that has $k_{\setS}$ copies of each set $\setS$ in $\collS$. 
Then the inequality \eqref{fii-rs} holds for the augmented collection
with the counts $r(i)$ and $r(\setS)$ appropriately modified. By considering 
arbitrary augmentations, one can obtain Theorem~\ref{thm:FII-gen} 
for fractional packings with rational coefficients. An approximation argument
yields the full version. This method of proof, although picturesque, is 
somewhat less transparent in the details.
\end{rmk}

\par\vspace{.15in}
\begin{rmk}
It is straightforward to extend Theorem~\ref{thm:FII-gen} 
and Corollary~\ref{cor:rs} to the multivariate case,
where $X_{i}$ are independent $\RL^{d}$-valued random vectors,
and $I(X)$ represents the trace of the Fisher information matrix
of $X$. Similarly, extending Theorem~\ref{thm:epi}, one obtains for 
independent $\RL^{d}$-valued random vectors $X_{1},\ldots, X_{n}$ 
with densities and finite covariance matrices that
\ben
e^{\frac{2H(X_{1}+\ldots+X_{n})}{d}} \geq \rth \sum_{\setS\in\collS} 
 e^{\frac{2H(\sum_{j\in\setS} X_{j})}{d}} ,
\een
which implies the monotonicity of entropy 
for standardized sums of $d$-dimensional random vectors.
We leave the details to the reader. 
\end{rmk}

\par\vspace{.15in}
\begin{rmk}
It is natural to speculate whether an analogous subset sum entropy 
power inequality holds with $1/r(\setS)$ inside the sum.  
For each $r$ between the minimum and the maximum of the 
$r(\setS)$, we can split $\collS$ into the sets 
$\collS_{r}= \{\setS\in\collS : r(\setS)=r\}$. Under an assumption that no one 
set $\setS$ in $\collS_{r}$ dominates, that is, that there is no 
$\setS^{*} \in \collS_{r}$ with 
$N(T^{\setS^{*}}) > \sum_{\setS\in\collS_{r}} N(T^{\setS})/r$, 
we are able to create suitable normals for perturbation of the 
Fisher information inequality and integrate (in the same manner as in the proof
of Proposition~\ref{prop:sums}) to obtain
\be\label{epi-rs}
N(T_n) \geq \sumS \frac{N(T^{\setS})}{r(\setS)}.
\ee
The quasibalanced case (in which $r(i)$ is the same for each $i$ in $\setS$) is an interesting 
special case.  Then the unions of sets in $\collS_{r}$ are disjoint for distinct 
values of $r$. So for quasibalanced collections the refined subset sum entropy power 
inequality \eqref{epi-rs} always holds by combining our observation above with the 
Shannon-Stam entropy power inequality.
\end{rmk}

\section{Concluding Remarks}
\label{sec:concl}

Our main contributions in this paper are rather general 
$\collS$-superadditivity inequalities for Fisher
information and entropy power that hold for arbitrary collections $\collS$,
and specialize to both the Shannon-Stam inequalities and the
inequalities of ABBN \cite{ABBN04:1}. In particular,
we prove all these inequalities transparently using only
simple projection facts, a variance drop lemma and classical
information-theoretic ideas.
A remarkable feature of our proofs is that their main ingredients
are rather well-known, although our generalizations of the variance drop lemma appear
to be new and are perhaps of independent interest.
Both our results as well as the proofs lend themselves to intuitive
statistical interpretations, several of which we have pointed out in the paper.
We now point to potential directions of application.

The inequalities of this paper are relevant to the study of
central limit theorems, especially for i.n.i.d. random variables.
Indeed, we demonstrated monotonicity on average properties
in such settings. Moreover, most approaches to entropic central limit
theorems involve a detailed quantification of the gaps associated with
monotonicity properties of Fisher information when the summands
are non-normal. Since the gap in our inequality is especially accessible due to our
use of a Pythagorean property of projections (see Remark~\ref{rmk:cltgap}), 
it could be of interest in obtaining transparent proofs
of entropic central limit theorems in i.n.i.d. settings, and perhaps rate
conclusions under less restrictive assumptions than those 
imposed in \cite{JB04} and \cite{ABBN04:2}.

The new Fisher information inequalities we present are also of 
interest, because of the relationship of inverse Fisher information to 
asymptotically efficient estimation. In this context, the subset sum
inequality can be interpreted as a comparison of an asymptotic mean
squared error achieved with use of all $X_{1}, \ldots, X_{n}$,
and the sum of the  mean squared errors achieved in distributed 
estimation by sensors that observe $(X_{i}, i\in\setS)$ for
$\setS\in\collS$. The parameter of interest can either be a location 
parameter, or (following \cite{JB04}) a natural parameter of exponential
families 
for which the minimal sufficient statistics are sums.
Furthermore, a non-asymptotic generalization of the new Fisher 
information inequalities holds (see \cite{BMKY07:pre} for details), 
which sheds light on minimax risks for estimation of a location parameter from sums.

Entropy inequalities involving subsets of random variables (although traditionally
not involving sums) have played an important role in understanding some problems of graph theory. 
Radhakrishnan \cite{Rad01} provides a nice survey, and some recent 
developments (including joint entropy inequalities analogous to the entropy power
inequalities in this paper) are discussed in \cite{MT07:isit}.
The appearance of fractional packings in the refined inequality we present
in Section~\ref{sec:ref} is particularly suggestive of further connections to be explored.

In multi-user information theory, subset sums of rates and 
information quantities involving subsets of random variables are critical in 
characterizing rate regions of certain source and channel coding
problems (e.g., $m$-user multiple access channels). Furthermore,
there is a long history of the use of the classical entropy power inequality
in the study of rate regions, see, e.g., Shannon \cite{Sha48},
Bergmans \cite{Ber74}, Ozarow \cite{Oza80}, Costa \cite{Cos85a}
and Oohama \cite{Ooh98}. For instance, the classical entropy power
inequality was a key tool in Ozarow's solution of the Gaussian multiple
description problem for two multiple descriptions, but seems to
have been inadequate for problems involving three or more descriptions
(see Wang and Viswanath \cite{WV05:pre} for a recent solution of one such problem
without using the entropy power inequality). It seems natural to expand the 
set of tools available for investigation in these contexts.

\appendices
\section{The Analysis of Variance Decomposition}

In order to prove the variance drop lemma, we use 
a decomposition of functions in $L^{2}(\RL^{n})$, which is nothing but the 
Analysis of Variance (ANOVA) decomposition of a statistic.
For any $j\in [n]$, $E_{j}\psi$ denotes the conditional expectation of $\psi$,
given all random variables other than $X_{j}$, i.e.,
\be\label{def:Ej}
E_{j}\psi(x_{1},\ldots,x_{n})=E[\psi(X_{1},\ldots,X_{n}) | X_{i}=x_{i}\quad \forall i\neq j ]
\ee
averages out the dependence on the $j$-th coordinate.

\par\vspace{.15in}
\noindent\begin{fact}[{\sc ANOVA Decomposition}]\label{f:dec}
Suppose $\psi:\RL^{n}\ra \RL$ satisfies $E\psi^{2}(X_{1},\ldots,X_{n})<\infty$,
i.e., $\psi\in L^{2}$, for independent random variables $X_{1},X_{2},\ldots,X_{n}$. 
For $\setT\subset [n]$, define the orthogonal linear subspaces
\be
\calH_{\setT} = \{ \psi\in L^{2}: E_{j}\psi=\psi 1_{\{j\notin \setT\}} \,\,\forall j\in[n]  \} 
\ee
of functions depending only on the variables indexed by $\setT$.
Then $L^{2}$ is the orthogonal direct sum  of 
this family of subspaces, i.e.,  any $\psi\in L^2$
can be written in the form
\be
\psi=\sum_{\setT\subset [n]} \Etb\psi,
\ee
where $\Etb\psi\in \calH_{\setT}$, and the subspaces 
$\calH_{\setT}$ \,(for $\setT\subset[n]$) are orthogonal to each other. 
\end{fact}
\par\vspace{.15in}

\begin{proof}
Let ${E}_{\setT}$ denote the integrating out of the variables
in $\setT$, so that $E_{j}=E_{\{j\}}$. Keeping in mind that the order of
integrating out independent variables does not matter
(i.e., the $E_{j}$ are commuting projection operators in $ L^2$), 
we can write
\be\begin{split}
\psi &= \prod_{j=1}^{n} [E_{j} + (I-E_{j})] \psi \\
&= \sum_{\setT\subset[n]} \prod_{j\notin \setT} E_{j} \prod_{j\in \setT} (I-E_{j}) \psi \\
&= \sum_{\setT\subset[n]} \Etb\psi,
\end{split}\ee
where 
\be
\Etb\psi \equiv {E}_{\setT^{c}} \prod_{j\in \setT} (I-E_{j}) \psi .
\ee
Note that if $j$ is in $\setT$, $E_{j}\Etb\psi=0$, being in the image of the operator
$E_{j}(I-E_{j})=0$. If $j$ is not in $\setT$, $\Etb\psi$ is already in the image of $E_{j}$,
and a further application of the projection $E_{j}$ has no effect. Thus 
$\Etb\psi$ is in $\calH_{\setT}$.

Finally we wish to show that the subspaces $\calH_{\setT}$ are
orthogonal. For any distinct sets $\setT_{1}$ and $\setT_{2}$ in $[n]$, 
there exists an index $j$ which is in one (say $\setT_{1}$), but not the other
(say $\setT_{2}$). Then, by definition, $\bar{E}_{\setT_{1}}\psi$ is contained in the image of $E_{j}$ 
and $\bar{E}_{\setT_{2}}\psi$ is contained in the image of $(I-E_{j})$. Hence 
$\bar{E}_{\setT_{1}}\psi$ is orthogonal to $\bar{E}_{\setT_{2}}\psi$.
\end{proof}
\par\vspace{.15in}

\noindent\begin{rmk}
In the language of ANOVA familiar to statisticians, 
when $\phi$ is the empty set, $\bar{E}_{\phi}\psi$
is the mean; $\bar{E}_{\{1\}}\psi, \bar{E}_{\{2\}}\psi, 
\ldots, \bar{E}_{\{n\}}\psi$ 
are the main effects; $\{\Etb\psi: |\setT|=2\}$ 
are the pairwise interactions, and so on. Fact~\ref{f:dec}
implies that for any subset $\setS\subset [n]$, the function
$\sum_{\{\setT: \setT\subset\setS\}} \Etb\psi $
is the best approximation (in mean square) to $\psi$ that depends
only on the collection $X_{\setS}$ of random variables. 
\end{rmk}
\par\vspace{.15in}
\noindent\begin{rmk}
The historical roots of this decomposition lie in the work of von Mises \cite{vonM47} and 
Hoeffding \cite{Hoe48}. For various interpretations, see Kurkjian and Zelen \cite{KZ62}, 
Jacobsen \cite{Jac68}, Rubin and Vitale \cite{RV80}, Efron and Stein \cite{ES81},
Karlin and Rinott \cite{KR82}, and Steele \cite{Ste86a}; these works include applications of such
decompositions to experimental design, linear models, $U$-statistics, and jackknife theory.
Takemura \cite{Tak83} describes a general unifying framework for ANOVA
decompositions.
\end{rmk}

\section*{Acknowledgment}

We are indebted to Prasad Tetali for several interesting conversations
and for the beneficial influence of Prasad's joint work with MM on 
entropy inequalities. We are also very grateful to the anonymous 
referees and the Associate Editor for careful readings of the 
paper, and their insightful comments. 


%

\end{document}